\documentclass[letter]{aa} % for the letters
\usepackage{graphicx,amssymb}
\usepackage{txfonts}
\def\gtsima{$\; \buildrel > \over \sim \;$}
\def\ltsima{$\; \buildrel < \over \sim \;$}
\def\prosima{$\; \buildrel \propto \over \sim \;$}
\def\gsim{\lower.5ex\hbox{\gtsima}}
\def\lsim{\lower.5ex\hbox{\ltsima}}
\def\simgt{\lower.5ex\hbox{\gtsima}}
\def\simlt{\lower.5ex\hbox{\ltsima}}
\def\simpr{\lower.5ex\hbox{\prosima}}

\def\h1{$h^{-1}$}
\def\eeq{\end{equation}}
\def\beq{\begin{equation}}
\def\24mu{24\,$\mu{\rm m}$}
\def\70mu{70\,$\mu{\rm m}$}
\def\8mu{8\,$\mu{\rm m}$}

\def\lya{Ly$\alpha$}

\begin{document}

   \title{The bending of the star-forming main sequence traces\\  the cold- to hot-accretion transition mass over $0<z<4$}

 %  \subtitle{I. Overviewing the $\kappa$-mechanism}

   \author{Emanuele Daddi
          \inst{1}
          \and          Ivan Delvecchio\inst{2}
          \and          Paola Dimauro\inst{3}
          \and          Benjamin Magnelli\inst{1}
          \and          Carlos Gomez-Guijarro\inst{1}
          \and   Rosemary Coogan\inst{1}
                    \and          David Elbaz\inst{1}
          \and 	  Boris~S. Kalita\inst{1}
                                        \and 	  Aurelien Le Bail\inst{1}
                    \and          R.~Michael Rich\inst{4}
          \and 	  Qing-hua Tan\inst{5,1}
          }
          
          \titlerunning{MS bending from  cold-accretion downfall}
          \authorrunning{E. Daddi et al.}

   \institute{CEA, IRFU, DAp, AIM, Universit\'e Paris-Saclay, Universit\'e Paris Cit\'e,  Sorbonne Paris Cit\'e, CNRS, 91191 Gif-sur-Yvette, France
          %    \email{edaddi@cea.fr}
         \and
         INAF-Osservatorio Astronomico di Brera, via Brera 28, 20121, Milano, Italy
         \and
         INAF - Osservatorio Astronomico di Roma, Via di Frascati 33, 00078, Monte Porzio Catone, Italy
         \and
         Department of Physics \& Astronomy, University of California Los Angeles, 430 Portola Plaza, Los Angeles, CA 90095, USA
         \and
         Purple Mountain Observatory, Chinese Academy of Sciences, 10 Yuanhua Road, Nanjing 210023, China
             }

  % \date{XXX}

% \abstract{}{}{}{}{}
% 5 {} token are mandatory
 
  \abstract
 {
 We analyse measurements of the evolving  stellar mass ($\mathcal{M}_0$ ) at which the bending of the star-forming main sequence (MS) occurs over $0<z<4$. We find $\mathcal{M}_0\approx10^{10}M_\odot$   over $0<z<1$, then $\mathcal{M}_0$  rises up to $\sim10^{11}M_\odot$ at $z=2$, and then stays flat or slowly increases towards higher redshifts. When converting $\mathcal{M}_0$  values into  hosting dark matter halo masses,  we show that this behaviour is remarkably consistent with the evolving cold- to hot-accretion transition mass, as predicted by theory and defined by the redshift-independent    $M_{\rm shock}$ at $z<1.4$ and by the rising $M_{\rm stream}$ at $z\gtrsim1.4$ (for which we propose a revision in agreement with latest simulations).  We hence argue that the MS bending is primarily due to the lessening of cold-accretion causing a reduction in available cold gas in galaxies and supports predictions of gas feeding theory.  In particular, the rapidly rising $\mathcal{M}_0$  with redshift at $z>1$ is confirming evidence for the cold-streams scenario. In this picture, a progressive fueling reduction rather than its sudden suppression in halos more massive than $M_{\rm shock}$/$M_{\rm stream}$  produces a nearly constant star-formation rate in galaxies with stellar masses larger than $\mathcal{M}_0$, and not their quenching, for which other physical processes are thus required. Compared to the knee $M^*$ in the stellar mass function of galaxies, $\mathcal{M}_0$  is  significantly lower  at $z<1.5$, and  higher at $z>2$, suggesting that the imprint of gas deprivation on the distribution of galaxy masses happened at early times ($z>1.5$--2). The typical mass at which galaxies inside the MS become bulge-dominated evolves differently from $\mathcal{M}_0$, consistent with the idea that bulge-formation is a distinct process from the  phasing-out of cold-accretion. 
 }

   \maketitle
%
%________________________________________________________________

\section{Introduction}

One of the key open issues in galaxy formation and evolution is understanding cold gas feeding of galaxies from the cosmic web sustaining their star-formation activity. 
Theory prescribes that cold gas  accretes freely onto dark matter halos with mass $M_{\rm DM}<M_{\rm shock}\approx10^{11.8}M_\odot$ at any redshift (Keres et al. 2005; Dekel \& Birnboim 2006; DB06 hereafter), and crucially also onto more massive halos at high redshift where cold-streams, collimated flows of cold accreting gas, can penetrate efficiently for halos with  $M_{\rm DM}<M_{\rm stream}(z)$ (DB06; Dekel et al. 2009). This $M_{\rm stream}(z)$ boundary is predicted to lie at $\sim10^{12.5}M_\odot$ at $z=2$, growing to $\sim10^{13.5}M_\odot$ at $z=3$. Observational evidence for cold accretion and for the cold-stream scenario is still scarce. 

Recently, Daddi et al. (2022; D22 hereafter) reported evidence for a progressive decrease in cold accretion for $M_{\rm DM}> M_{\rm stream}$ onto massive groups and clusters over $2<z<3.3$. They inferred that the fraction of baryonic accretion rate (BAR) remaining cold scales down like $(M_{\rm stream}/M_{\rm DM})^\alpha$, with slope $\alpha\sim1$ and no apparent discontinuity up to two~dex above $M_{\rm stream}$.  This result was primarily based on the extended \lya\ emission luminosities, but  slopes consistent with $\alpha\sim1$ appeared to characterise also the star-formation rate (SFR) and bolometric active galactic nucleus (AGN) luminosities integrated over the massive halos. A modulation with slope $\alpha\sim1$ implies that any quantity scaling linearly with cold accretion, and thus rising (at fixed redshift) proportionally to $M_{\rm DM}$ for $M_{\rm DM}<M_{\rm stream}\ ({\rm or\ }M_{\rm shock})$, would remain roughly constant at larger $M_{\rm DM}>M_{\rm stream}$ (see Eqs.~5~and~6 in D22, for \lya). 

It is intriguing that a similar mass dependence has been already recognised for the bending of the star-forming Main Sequence (MS), where SFR rises nearly linearly until a certain bending stellar mass ($\mathcal{M}_0$ ), and stays constant at larger masses (e.g., { Ilbert et al. 2015}; Lee et al. 2015; Schreiber et al. 2015; 2016; Popesso et al. 2019). Accretion rates of cold gas should have a direct impact on the SFRs, hence it is quite natural that the distribution of SFRs in galaxies (and the MS) is an important observable where variations in accretion modes could be detected.
The bending mass at $z\sim0$ has been already noticed to be similar to the  $M_{\rm shock}$ boundary (Popesso et al. 2019), once the average  stellar to halo mass relation (SHMR) is considered (Behroozi et al. 2013). Bending gets progressively weaker over $10^{10}M_\odot<M_*<10^{11}M_\odot$ at $z>1$ (Schreiber et al. 2015; Delvecchio et al. 2021), suggesting that $\mathcal{M}_0$  increases strongly with redshift, qualitatively similar to the rise of the $M_{\rm stream}$ boundary. The analogies are so strong to warrant a quantitative examination. 

In this letter we study the relation between $\mathcal{M}_0$  and $M_{\rm shock}/M_{\rm stream}$ as a function of redshift, showing that they are essentially the same mass scale over $0<z<4$, and discuss the implications of this coincidence. We adopt concordance cosmology (0.3; 0.7; 70) and a Chabrier IMF.

\begin{figure}[ht]
\centering
\includegraphics[width=9cm,angle=0]{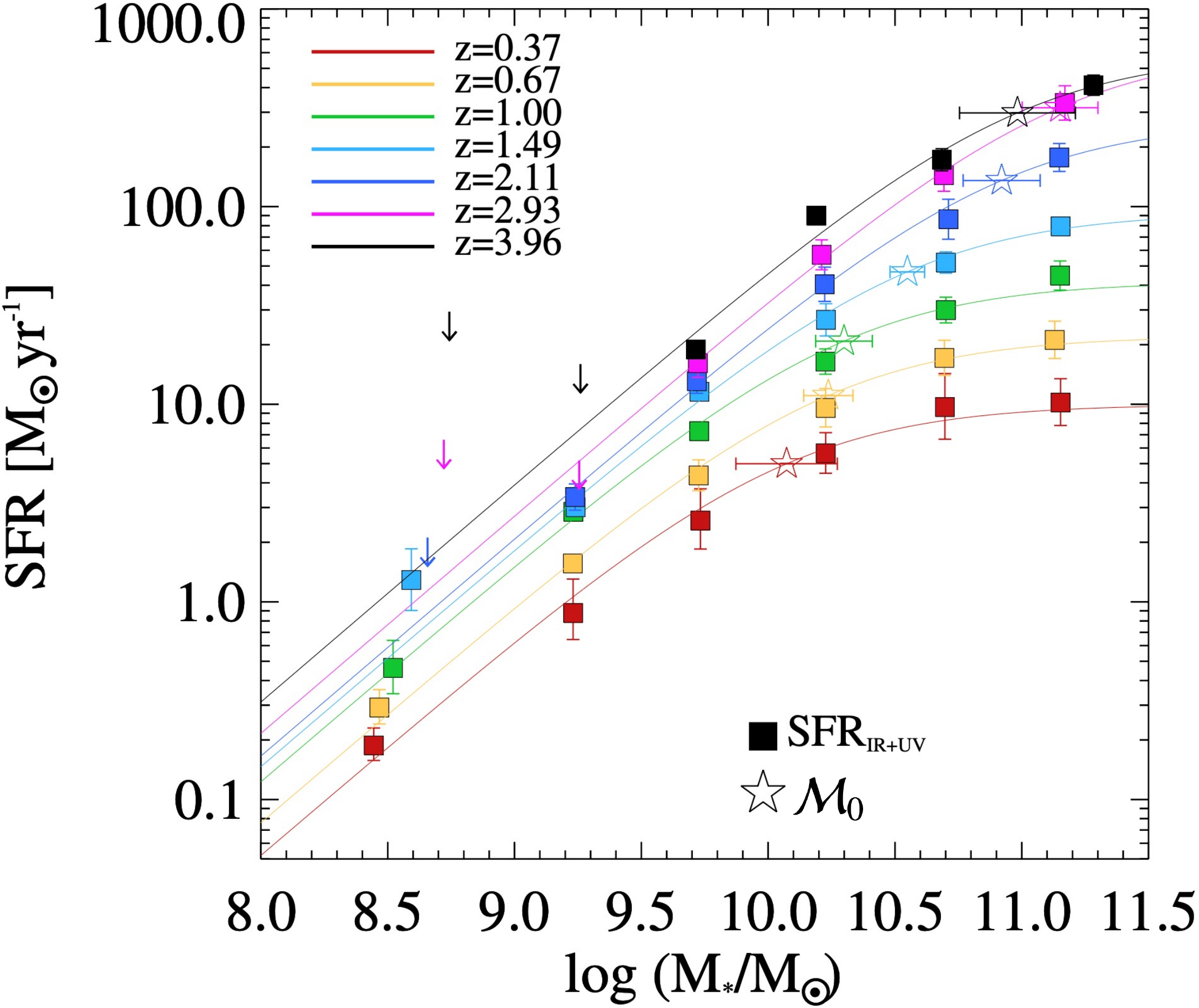}
\caption{The star-forming Main Sequence derived in redshift bins over $0.4<z<4.0$ (squares), adapted from Delvecchio et al. (2021). Solid lines show  fits of Eq.~\ref{eq:M0} to the data. The bending stellar mass $\mathcal{M}_0$  (see text for details) is shown as an empty star for each redshift bin (with its error). Notice how it rapidly increases from low- to high-redshifts.
}
\label{fig:ivan}
\end{figure}

\section{Quantifying the redshift evolution of the bending of the star-forming Main Sequence}

\subsection{Bending formalism}

The parametrisation for describing the MS (SFR vs. the stellar mass $M_*$)   with its bending is adopted\footnote{With the purely estethical difference that we express parameters in linear instead of log space.} from Lee et al. 2015 (L15 hereafter), that we rewrite as:

\beq
\frac{\rm SFR}{\rm SFR0} = \frac{1}{1+(\mathcal{M}_0/M_*)^\gamma}
\label{eq:M0}
\eeq

with  $\mathcal{M}_0$  being the {\em bending mass}, SFR0  the SFR saturation limit  for $M_*>>\mathcal{M}_0$, and $\gamma$ the MS slope  in the limit of $M_*<<\mathcal{M}_0$.  Note that Eq.~\ref{eq:M0} implies SFR($\mathcal{M}_0$ )/SFR0~$=0.5$ for any $\gamma$, hence the SFR only marginally further rises by $\times2$ beyond $\mathcal{M}_0$. 

 \begin{table}
\centering
\caption{Fitting of the star-forming MS with bending, in different redshift bins. 
\label{table1}}
\begin{tabular}{lrrr}
 \hline
z & log $\mathcal{M}_0$  & log SFR0  &  $\gamma$  \\
 &  $M_\odot$ &  $M_\odot$~yr$^{-1}$ &  \\ 
% & & &(1)  & (2) & (3) & \\
 \hline
 \multicolumn{4}{l}{Lee et al. 2015}\\
0.36 & $10.03 \pm 0.14$ & $0.80 \pm 0.07$ & $0.92 \pm 0.06$  \\
0.55 & $9.82 \pm 0.11$ & $0.99 \pm 0.05$ & $1.13 \pm 0.11$  \\
0.70 & $9.93 \pm 0.11$ & $1.23 \pm 0.06$ & $1.11 \pm 0.09$  \\
0.85 & $9.96 \pm 0.09$ & $1.35 \pm 0.05$ & $1.28 \pm 0.12$  \\
0.99 & $10.10 \pm 0.10$ & $1.53 \pm 0.06$ & $1.26 \pm 0.11$  \\
1.19 & $10.31 \pm 0.15$ & $1.72 \pm 0.08$ & $1.07 \pm 0.10$ \\
\hline
 \multicolumn{4}{l}{Delvecchio et al. 2021}\\
 0.37 & $10.06 \pm 0.18$ & $1.02 \pm 0.11$ & 1.1  \\
0.67 & $10.25 \pm 0.09$ & $1.34 \pm 0.09$ & 1.1  \\
1.00 & $10.30 \pm 0.11$ & $1.62 \pm 0.10$ & 1.1  \\
1.49 & $10.56 \pm 0.07$ & $1.98 \pm 0.06$ & 1.1 \\
2.11 & $10.92 \pm 0.14$ & $2.43 \pm 0.13$ & 1.1  \\
2.93 & $11.15 \pm 0.16$ & $2.81 \pm 0.15$ & 1.1  \\
3.96 & $11.00 \pm 0.21$ & $2.79 \pm 0.19$ & 1.1 \\
\hline
\end{tabular}
{\footnotesize \ \ \ \ \ \ \ \ \ \ \ Notes: Lee et al. (2015) measurements were presented originally in their work, we have only scaled-up the uncertainties (see text). The parameters for  Delvecchio et al. (2021)  are derived in this work. }
\vspace{0.5truecm}
\end{table}

There are strong analogies between Eq.~\ref{eq:M0} and the formalism from D22 (their Eqs. 1--4) describing quantities scaling proportionally to cold accretion.  While in D22 the equations relate to $M_{\rm DM}$, and here to $M_*$,  stellar and dark matter mass scales are tightly related on average (e.g., Behroozi et al. 2013). Also, the D22 formalism describes a stepwise behaviour across the $M_{\rm shock}/M_{\rm stream}$ boundaries (Eq.~3 in D22) while here we have a continuous function (Eq.~1), but the two can be shown to be fully consistent. We built a simple toy model with the expected behaviour from Eq.~3 in D22, adding appropriate noise in SFR as well as in the SHMR. We find that $\mathcal{M}_0$  as defined in Eq.~1 is an unbiased estimator of the stepwise boundary (i.e., $M_{\rm shock}/M_{\rm stream}$) in the presence of noise, to better than 0.02~dex. 

The fact that Eq.1 converges to a constant SFR for $M_*>>\mathcal{M}_0$  is equivalent to  $\alpha_{\rm SFR}\sim1$ in D22 (and indeed the key result from D22 was to constrain $\alpha_{Ly\alpha}\sim1$). The D22 formalism adopts in practice $\gamma=1.15$ that is inherited by the predicted $M_{\rm DM}$ dependence of the BAR onto DM halos (Goerdt et al. 2009; Genel et al. 2008; Dekel et al. 2013) in the cold-stream or cold-accretion regimes. Indeed, as we will see in the following, the MS also implies $\gamma\sim1.1$. This analogy between the MS slope and BAR (before bending happens) has already been discussed by Dekel et al. (2013). All of this suggests that to some extent $\mathcal{M}_0$  might be equivalent to $M_{\rm shock}/M_{\rm stream}$, which we will demonstrate empirically in the following by deriving the redshift evolution of $\mathcal{M}_0$  since $z=4$ and comparing it to the $M_{\rm shock}/M_{\rm stream}$ masses over the same range. 

\subsection{Redshift evolution of bending}

Obtaining reliable fits of the parameters from Eq.~\ref{eq:M0} requires:  (1) a deep sample, reaching much fainter than $\mathcal{M}_0$  (at any redshift, which is made easier by the implied redshift evolution of $\mathcal{M}_0$). Also, (2)  large statistics at the high $M_*$ end are required, in order to derive SFR0, which demands large sky area; and (3) the use of  consistent SFR indicators across the entire dynamic range. In this work we do not obtain new MS determinations, something already done in countless literature works (see Speagle et al. 2014 for a review). We adopt average SFR vs stellar mass ($M_*$) measurements from state of the art works (L15 and Delvecchio et al. 2021; D21 hereafter) spanning (when combined) $0<z<4$ with large statistics and reaching down to very low stellar masses which is essential for this analysis. These two works fulfill the mentioned requirements and are both based in the COSMOS field. 

\begin{figure*}[ht]
\centering
\includegraphics[width=18.2cm,angle=0]{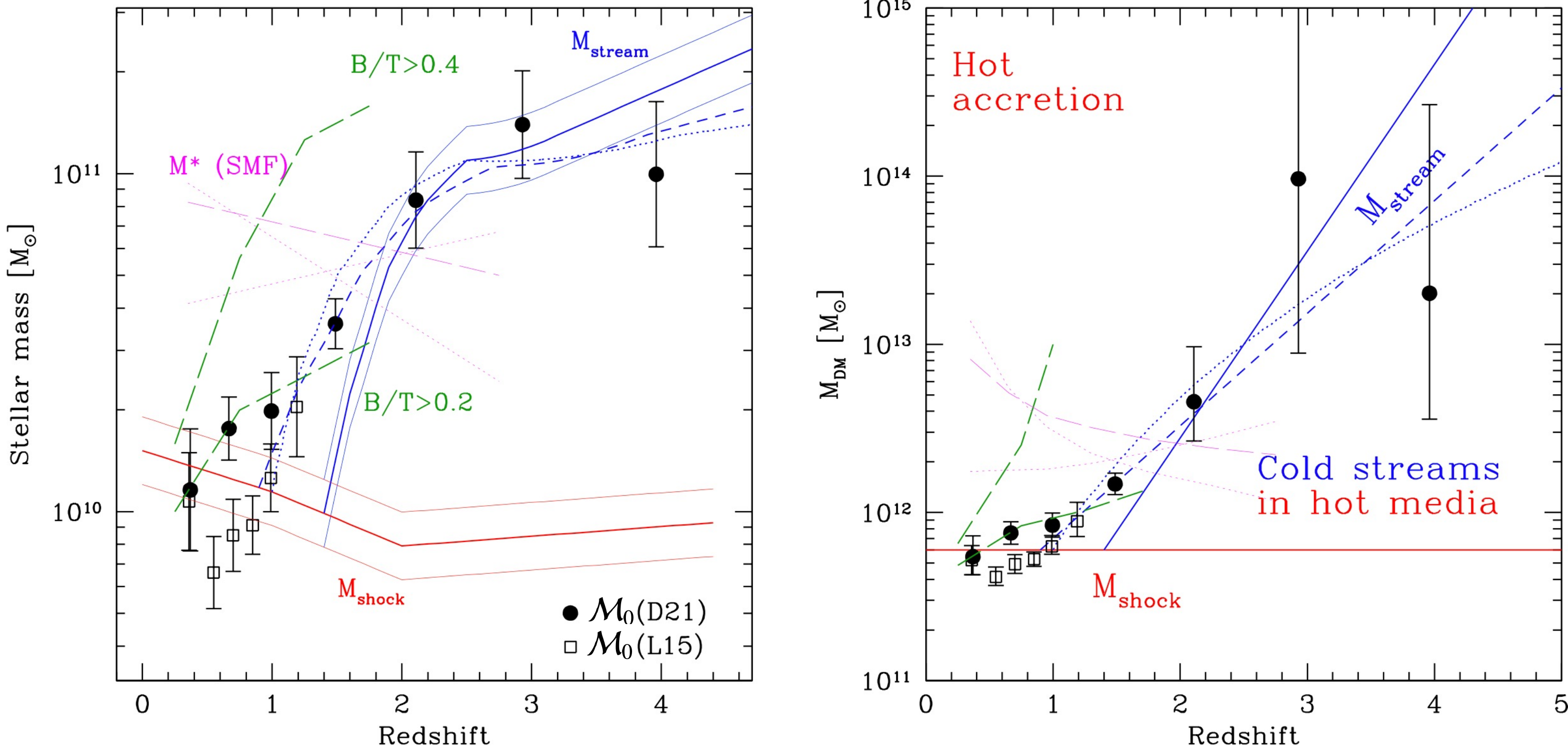}
\caption{Measurements of $\mathcal{M}_0$  are shown in both panels from L15 (empty squares) and D21 (filled circles), together with the $M_{\rm shock}$ and $M_{\rm stream}$ boundaries (DB06,  solid blue line; from our fit to the $\mathcal{M}_0$  data as in Eq.~2, dashed blue line; based on Mandelker et al. 2020, dotted blue line), and  stellar mass function' (SMF)  M* values   (Ilbert et al. 2013; linear fit to their redshift trends; solid for the total, dotted for those of quiescent/SF galaxies that are decreasing/increasing with redshift, respectively). The masses at which the average bulge/total (B/T) ratio in MS galaxies rises above 0.2 and 0.4 are based on Dimauro et al. (2022; green long-dashed). Measurements and relations are converted from stellar to halo masses (and vice-versa) using the SHMRs from Behroozi et al. (2013). {  In doing this we ignore the possible difference between the direct and inverse SHMR, which could have some impact at the highest masses (O. Ginzburg et al., in preparation). The effect of varying by $\pm0.1$~dex the SHMR is shown by the thinner version of the DB06 tracks. This is larger than the statistical uncertainties of the best measurements (e.g., Shuntov et al. 2022) on average, but appropriate to encompass systematics from different methods (e.g., Behroozi et al. 2019).}
}
\label{fig:mass}
\end{figure*}

L15 measured the MS over 6 redshift bins with average $z=0.36$ to 1.19,  using $\sim62,000$ SF galaxies selected from Ilbert et al. (2013) using the NRK color method. SFRs are derived using a ladder approach, including UV, mid-IR and far-IR measurements. 
Their MS is obtained using a very finely binned grid to which they fitted Eq.~\ref{eq:M0}, and report the 3 free parameters with  formal uncertainties from the fit. From their Fig.~6 one can see that their $\mathcal{M}_0$  values display a non-monotonous behaviour that is seemingly significant with respect to their reported error bars. E.g., $\mathcal{M}_0$  significantly decreases from $z=0.34$ to 0.51, and jumps up again at $z=0.70$--0.99. We take this as evidence of underestimation of the  uncertainties, e.g. due to degeneracies between parameters, and systematics. We find that scaling up the uncertainties by a factor of 3.4 is sufficient to remove the anomalous redshift fluctuations.  A posteriori based on the minimum reduced $\chi^2$ of the fit to all datasets, and also in comparison with the D21 dataset, we confirm that this scaling is indeed required. We apply this to all L15 parameter uncertainties. The L15 results with the proposed error rescaling are summarised in Table~1. 

D21 measured the MS over 7 redshift bins with average $z=0.37$ to 3.96, using $\sim400,000$ SF galaxies from Laigle et al. (2016),   selected using the NRJ color method (basically identical to the NRK but improving over it for the highest redshift range, implying full consistency with the L15 sample). For each redshift bin they defined large, 0.5~dex-wide stellar mass bins (except for the lowest and highest mass bins that are 1~dex wide) and measured the average SFR via stacking in multi-band far-IR datasets, including Herschel, Spitzer, SCUBA and AzTEC, and adding the contribution of un-obscured UV. Because their binning is wide,  to reduce the noise in the measurements we fixed the value of $\gamma=1.1$,  the average slope obtained by L15  using a much finer grid. Using free $\gamma$ would not alter the conclusion of this work (and on average would still give a consistent $\gamma\sim1.1$). We thus fitted Eq.~\ref{eq:M0} to the D21 data (Fig.~\ref{fig:ivan})  and estimated uncertainties in the best fitting parameters from Monte Carlo simulations. The measurements are in Table~1.  

The $\mathcal{M}_0$  measurements as a function of redshift are shown  in Fig.~\ref{fig:mass}-left. Derivations from L15 and D21 are in reasonably good agreement over $0.4<z<1.2$ where they define an average at the level of  $\mathcal{M}_0\sim10^{10}M_\odot$. The scatter between the two datasets allows us to gauge the underlying systematics. 
Mainly based on the D21 values, but also supported by the highest redshift point from L15 ($z=1.2$), we see that $\mathcal{M}_0$  rapidly increases above $z=1$, to reach $\mathcal{M}_0\sim10^{11}M_\odot$ at $z\gtrsim2$. 

\subsection{Bending in halo mass space}

We convert each $\mathcal{M}_0$  measurement into the corresponding average $M_{\rm DM}$ value using the redshift-dependent relations from Behroozi et al. (2013; their Fig.~7-left). Such relations are defined for central galaxies, or recently accreted satellites, but can be applied safely here because the vast majority of galaxies over the stellar mass ranges considered here are centrals (Popesso et al. 2019; McCracken et al. 2015).
Fig.~\ref{fig:mass}-right  shows the evolution of $\mathcal{M}_0$  when expressed in terms of average hosting DM halo mass. At $z<1$ the typical $\mathcal{M}_0$  corresponds to $\sim4$--$8\times10^{11}M_\odot$, which is quite consistent with $M_{\rm shock}$. Beyond $z\sim1$ we see a rise until reaching $\mathcal{M}_0\sim10^{13-14}M_\odot$ over $2<z<4$. This rise follows very closely the redshift dependence of $M_{\rm stream}$ and is quite consistent with its definition from DB06. The $\mathcal{M}_0$  values expressed in terms of $M_{\rm DM}$ display large uncertainties at high masses/redshifts, due to the flattening of the $M_*/M_{\rm DM}$ relation at $M_*\gtrsim10^{11}M_\odot$ (Behroozi et al. 2013), so that even relatively small uncertainties in $M_*$ convert into large uncertainties in $M_{\rm DM}$. To cope with this issue, in Fig.~\ref{fig:mass}-left we show the $M_{\rm shock}$ and  $M_{\rm stream}$ tracks converted from dark matter to stellar masses in the same way.  $\mathcal{M}_0$  measurements are in good agreement with  $M_{\rm stream}$ expectations at $z\gtrsim2$. 

\section{Refining the $M_{\rm stream}$ boundary}

From Fig.~\ref{fig:mass} it is quite apparent that $\mathcal{M}_0$  and $M_{\rm shock}/M_{\rm stream}$ are basically the same scale. However, there is a small discrepancy in the range $1<z<1.4$, due to the fact that $\mathcal{M}_0$  rises up earlier than $M_{\rm stream}$ as originally predicted in DB06. As discussed at length in Dekel et al. (2009), the $M_{\rm stream}$ boundary was defined using  ad-hoc assumptions, and its location would benefit from being refined by observations.  

We use the evolving $\mathcal{M}_0$  data to find the best fitting $M_{\rm stream}(z)$ relation, allowing its low-redshift intercept  $z_{\rm stream, min}$ and its slope to be free parameters (while keeping $M_{\rm shock}$ fixed at the DB06 value for $z<z_{\rm stream, min}$). The best-fitting relation is:

\beq
{\rm log}\ M_{\rm stream} \simeq {\rm log}\ M_{\rm shock} +(0.67\pm0.15)\times(z-0.9\pm0.1) 
\label{eq:mstream}
\eeq

and it is shown in Fig.~\ref{fig:mass}  as a dashed blue line. The best fit has $\chi^2_{\rm min}\sim17.5$ with 11 d.o.f, which has a 10\% probability to occur by chance (hence acceptable), showing that the error bars from our measurements are not severely underestimated. 
The two free parameters are somewhat correlated, in the sense that a higher $z_{\rm stream, min}$ corresponds to a steeper relation. 
This relation has a lower $z_{\rm stream, min}$ (0.9 vs 1.4) and a flatter slope (0.67 vs 1.11) with respect to the one from DB06 (see Eq.~2 in D22), which is a  poorer representation of the $\mathcal{M}_0$  data. %($\Delta\chi^2\sim37)$. 
{ Evidence for a flatter $M_{\rm stream}$ slope is even stronger when adopting the latest SHMR derivations at $z>2$ from Shuntov et al. 2022.}

Quite interestingly, if we were to adopt Eq.~\ref{eq:mstream} as a definition of $M_{\rm stream}$ and repeat the analysis from D22 based on observations of 9 massive groups and clusters of galaxies at $2<z<3.3$, we find overall improvement  in the significance of the D22 results. For example, the \lya\ luminosities would provide $\alpha_{\rm Ly\alpha} = 1.05\pm0.18$ (5.8$\sigma$, vs 5.0$\sigma$ in D22) with a scatter of 0.28~dex (0.30~dex in D22). Similarly, the correlation with the integrated SFRs would reach  2.9$\sigma$ (vs 2.6$\sigma$ in D22), with a scatter of 0.43~dex (vs. 0.45~dex). There is thus encouraging and independent observational support that this revised relation might be a change in the right direction. 

%\subsection{A flatter $M_{\rm stream}(z)$ is consistent with state of the art simulations}

Present-day numerical simulations and analytical work suggest that a crucial process in the cold-stream feeding of cold gas to halos/galaxies  is  the cooling of hot halo gas through the radiative turbulent mixing layer that forms at the boundary between the initially cold stream and the initially hot background circum-galactic medium (CGM; Mandelker et al. 2020a; Gronke \& Oh 2020, Fielding et al 2020). The survival of the streams to hydrodynamical instabilities in the CGM until reaching the central galaxy  requires that the cooling time in the mixing layer is significantly shorter than the disruption time, especially in highly turbulent environments such as the CGM (Gronke et al. 2022). This can equivalently be expressed as a requirement that the stream radius ($R_{\rm s}$) is  larger than the critical radius ($R_{\rm crit}$) by a substantial factor (Gronke \& Oh 2020; Kanjilal et al. 2021; Mandelker et al. 2020b). Defining $M_{\rm stream}(z)$ as the locus where $R_{\rm s}/R_{\rm crit}=20$ from the fiducial model in Mandelker et al. (2020b; see their Fig.~2 where this ratio is expressed as a function of $M_{\rm DM}$ and $z$) returns the dotted blue line shown in Fig.~2, which is in striking agreement with the $\mathcal{M}_0(z)$  data and our $M_{\rm stream}(z)$ refinement (Eq.~2). 

\section{Discussion}

The $M_{\rm shock}/M_{\rm stream}$ boundaries mark, as a function of redshift, the critical masses when cold-accretion is expected to subside and turn into hot-accretion, based on theory predictions. Fig.~\ref{fig:mass}  shows that these mass scales are in good agreement with $\mathcal{M}_0(z)$, the mass at which the MS is bending, over $0<z<4$.\footnote{ Recently, Popesso et al. (2022) reached, independently, a consistent conclusion.} This is unlikely to happen by chance, given the complex redshift dependence. Also,  there is a fairly reasonable physical explanation for this coincidence: when cold-accretion starts to be reduced, becoming an increasingly smaller fraction of the total baryonic accretion, less fuel is available to galaxies to form stars, hence the SFRs of such massive galaxies deviate from the SFR-$M_*$ trend defined by lower-mass galaxies, as advocated in gas-regulator models (e.g., Bouch\`e et al. 2010; Lilly et al. 2013). The lower gas fraction of galaxies above $\mathcal{M}_0$  is implied by the constancy of the star-formation efficiency versus mass across the $M_{\rm shock}/M_{\rm stream}$ boundary (e.g., Wang et al. 2022).
%The same overall scenario had already been discussed for the extended \lya\ emission from massive halos across the $M_{\rm stream}$ boundary from D22, where \lya\ luminosity is also assumed to be a probe of cold-accretion rate. In addition, D22 also found evidence for a similar trend for the integrated SFR over massive halos, albeit at moderate significance. 
The main consequence of this coincidence between $\mathcal{M}_0$  and $M_{\rm shock}/M_{\rm stream}$ is thus that it appears inevitable to conclude that the bending of the MS, through cosmic time, is primarily due to the phasing out of cold accretion.
%\subsection{Cold accretion downfall does not produce quenching}
 Additionally, this allows us to go a step further, and notice that the physical consequence of the phasing-out of cold accretion is to produce a flat SFR-$M_*$ relation, i.e. reaching constant SFR vs $M_*$ as an asymptotic value (SFR0). This is equivalent to say, formally, that $\alpha\sim1$  (see intro; D22 -- { this will include residual cold accretion and contributions from hot gas cooling}).  A constant SFR vs $M_*$, and at fairly high values, is still quite some relevant amount of activity, at odds with the often held conception that the shutting down of cold-accretion beyond the $M_{\rm shock}/M_{\rm stream}$  boundary  induces {\em quenching}.  We recall that quenched galaxies are defined as those with very low amounts of residual SFR, e.g. much lower than that of the MS (including its bending). 

A slope $\alpha\sim1$ implies, for example, that a halo with mass 10$\times$ above the boundary has a cold share of accretion of 10\% of the total (Eq.~3 in D22). However, such a halo has a total accretion  10$\times$ larger than a halo  at the boundary (Eq.~1 in D22). Ultimately the cold accretion rate remains constant at any halo mass above the boundary. Hence, even when cold accretion is strongly reduced in its effective fraction, in the range of masses  that we have been probing it is not reduced in absolute scale, preventing galaxies from becoming quenched. The obvious consequence of this is that there is not such a thing as {\em starvation} from cold gas accretion discontinuation, at least not in the typical ranges that we have probed (up to 1~dex above $\mathcal{M}_0$  at $z<1$ from this work, and up to 2~dex above $M_{\rm stream}$ at $z\sim2$ from \lya\ in D22); cold gas shut down is not sufficient  for galaxies to starve to death. Physical processes other than the shutting down of cold-accretion are required to quench galaxies. This might well be mergers (e.g., Puglisi et al. 2021), AGN (e.g., Brusa et al. 2018) or instability driven (e.g., Kalita et al. 2022).

We exclude the possibility that cold accretion contraction in the hot-accretion regime induces lots of real quenching, but those galaxies escape from our sample due to selection effects. While this might be suspected, as by construction we study only SF galaxies, at the masses where $\mathcal{M}_0$  occurs the SF galaxies are much more numerous than quiescent ones, over the whole redshift range studied (Ilbert et al. 2013; their Fig.6). 

%\subsection{Bending mass and the galaxy mass function of SF and quenched galaxies}

In order to further gauge the significance of our results, it is worth comparing $\mathcal{M}_0$  to the  characteristic `knee' masses (M*) in the galaxy stellar mass functions (SMFs), representing the mass at which the differential mass (and SFR) contribution peaks, and the mass above which galaxies become exponentially rarer. These are shown as magenta lines in Fig.~\ref{fig:mass} (including the total SMF, and SMFs for quiescent and SF galaxies). At $z<1.5$, M* in the SMF is always much more massive than the bending mass, while at $z\gtrsim2$ it is less massive.  Hence, the cold-accretion to hot-accretion transition does not appear to relate in any simple way to M*  in the  SMF of galaxies, and the definition of {\em typical} galaxy stellar masses.  The M* for SF galaxies though does  decrease mildly with cosmic time, possibly influenced by the similar decrease of $\mathcal{M}_0$.  As  stars in quenched galaxies observed at $z\sim0$--1 are very old, formed at early times, this supports the idea that the imprint of the downfall of gas accretion into the galaxies SMF also happened at early times ($z>1.5$--2 -- this is because the lessening of accretion will reduce the ability of galaxies to  become much more massive). The M* for quenched galaxies runs almost perpendicular to $\mathcal{M}_0$, increasing with cosmic time towards low-redshift, emphasising furthermore the poor causal connection between the build-up of quenched galaxies and cold-accretion presence or suppression thereof. This is  consistent with the recent finding (Kalita et al. 2021) of an extremely old quenched galaxy in the cold gas rich RO-1001 environment (Daddi et al. 2021), and in general with the presence of quenched galaxies at $\gtrsim3$--4 (Valentino et al. 2020; D'Eugenio et al. 2020; 2021; Forrest et al. 2020).  

%\subsection{Cold accretion downfall unrelated to the rise of bulges}

It has been often remarked that galaxies more massive than the MS bending are increasingly hosting more massive bulges, to the extent that considering only the disk-like components and ascribing to them the SFR, one gets closer to a MS rising trend (Abramson et al. 2014; Mancini et al. 2019; Dimauro et al. 2022). It is worth reconsidering the matter here, in light of our systematic discussion of the evolution of $\mathcal{M}_0$. We use the results of Dimauro et al. (2018; 2019; 2022), that performed bulge-disk decomposition in $0<z<2$ galaxies in CANDELS, using a machine learning approach for identifying bulge+disk systems. Limiting the results only to color-selected SF galaxies as in this work, the stellar masses along the MS in which the average bulge/total (B/T) mass fraction in galaxies exceeds 0.2 and 0.4 can be derived (Fig.~\ref{fig:mass}).  The $B/T=0.4$ threshold is generally much higher than $\mathcal{M}_0$, although getting closer to it towards $z=0$. Clearly most of the mass in MS galaxies is contained in the disk, even well beyond $\mathcal{M}_0$, but the bulge fraction keeps increasing with time, probably due to the hierarchical assembly of the bulges, and possibly rejuvenation (e.g., Mancini et al. 2019). The $B/T=0.2$ threshold is closer to $\mathcal{M}_0$ over $0<z<2$. It still has though a distinct rise with redshift that is incompatible with the  constant $M_{\rm shock}$ boundary. It appears thus that the growth of the bulges has little in common with the phasing out of cold accretion, as might be reasonably expected also because the high Sersic indices of bulges require more than simple removal of gas infall, but likely mergers. This conclusion is supported by the analysis in Freundlich et al. (2019) where no connection between gas scaling relations and morphology was detected.  It is tantalising though that the $B/T=0.4$ trend mimics quite closely the $M_{\rm stream}$ trend but shifted roughly by $\Delta z\sim1$ to later times. This shift might track the required timescale between the formation of bulge stars and the assembly and quenching of those same bulges.  

%\subsection{Relation to $M^{\rm peak}_{\rm DM}$}

{ The location of the maximum in the stellar to halo mass relation ($M^{\rm peak}_{\rm DM}$; e.g., Behroozi et al. 2013) can also be affected by the accretion history of galaxies. Best current determinations show that this is consistently more massive than $\mathcal{M}_0$ by 0.3-0.5dex over $0.3<z<1.5$ where it is most accurately determined (e.g., Legrand et al. 2019; Shuntov et al. 2022), but possibly consistent at $z>2$. More generally, our formalism with $\alpha=1$ implies that beyond $\mathcal{M}_0$ one could expect DM vs $M_*$ differential growth rates producing asymptotically $M_*/M_{\rm DM}\propto {\rm log}\ M_{\rm DM}/M_{\rm DM}$ which is similar to the observed SHMRs at the highest masses $M_{\rm DM}>10^{13}M_\odot$. This is still consistent with the idea that feedback from star-formation has a dominant role in shaping the $M_*/M_{\rm DM}$ relation, particularly at intermediate/low masses ($M^{\rm peak}_{\rm DM}$) and at later epochs. }

%\subsection{Evolution of bending as confirming evidence for the cold-streams scenario}

The slope of the MS has been generally reported to be in the range of 0.5--0.9 (Speagle et al. 2014 for a review), shallower than the dependence of BAR on the mass which reads 1.15 (e.g., Dekel et al. 2013). However the observational measurements are 
affected  by the bending of the MS. The limiting slope at low masses would be a better comparison ground, and this is indeed typically $>1$, on average 1.1 in L15, remarkably close to the expected theory value. In addition, the redshift evolution rate in the normalisation of the MS has been estimated as going in the range $(1+z)^{2.8-3.5}$ (Speagle et al. 2014 for a review), much faster than the exponent of 2.25--2.5 expected from theory. Again, this is typically derived at $M_*\sim5\times10^{10}M_\odot$, where all the values at $z<2$ are strongly affected by bending, producing a spuriously fast evolution. When considering the MS bending, the overall behaviour of the MS seems to be in even much better agreement with theory than what was realised so far. The rise of $\mathcal{M}_0$ aligned to that of $M_{\rm stream}$ is also further confirmation of the cold-stream theory, as discussed in this work.
\smallskip

To conclude, together with the results from D22, the evidence is accumulating that cold streams (or their modulation) have a measurable impact on galaxy formation and evolution. Systematically investigating the physical properties of galaxies above and below $\mathcal{M}_0(z)$  will further enlighten the processes by which accretion from the cosmic web affects galaxies. 
\smallskip

%Limitations due to M*/Mh conversion, particularly at highz, have an impact ?\\

%bending and rejuvenation, in Mancini+2019 we said it can largely explain bending\\

%bending due to AGN ?\\

\begin{acknowledgements}
     We thank Nir Mandelker for providing us with a current theory-based derivation of $M_{\rm stream}$ based on his recent work and, together with Avishai Dekel, for helpful discussions. The anonymous referee is acknowledged for a constructive report. 
\end{acknowledgements}

\end{document}